\documentclass[twocolumn,prl]{revtex4} 
\usepackage{graphicx} 
\usepackage{bm} 
\usepackage{amsmath} 
\usepackage{color}

\newcommand{\Ket}[1]{\vert  #1 \rangle}

\newcommand{\be}{\begin{equation}} 
\newcommand{\ee}{\end{equation}} 
\newcommand{\bea}{\begin{eqnarray}} 
\newcommand{\eea}{\end{eqnarray}}

\renewcommand{\phi}{\varphi} 
\renewcommand{\epsilon}{\varepsilon}

\begin{document} 
 
\title{Self-sustaining dynamical nuclear polarization oscillations in quantum dots}

\author{M. S. Rudner$^{1,2,3}$ and L. S. Levitov$^{4}$}
\affiliation{ 
$^{1}$ The Niels Bohr International Academy, Blegdamsvej 17, DK-2100 Copenhagen \O, Denmark\\
$^{2}$ Department of Physics, The Ohio State University, 191 W. Woodruff Ave., Columbus, OH 43210\\
$^{3}$ Institute for Quantum Optics and Quantum Information of the Austrian Academy of Sciences, A-6020 Innsbruck, Austria\\
$^{4}$ Department of Physics, Massachusetts Institute of Technology, 77 Massachusetts Ave., Cambridge, MA 02139
} 
 
 
\begin{abstract} 

  Early experiments on spin-blockaded double quantum dots revealed surprising robust, large-amplitude current oscillations in the presence of a static (dc) source-drain bias [see e.g.~K. Ono, S. Tarucha, Phys. Rev. Lett. {\bf 92}, 256803 (2004)].
Experimental evidence strongly indicates that dynamical nuclear polarization plays a central role, 
but the mechanism has remained a mystery.
Here we introduce a minimal albeit realistic model of coupled electron and nuclear spin dynamics which supports robust self-sustained oscillations.
Our mechanism relies on a nuclear-spin analog of the tunneling magnetoresistance phenomenon (spin-dependent tunneling rates in the presence of an inhomogeneous Overhauser field) and nuclear spin diffusion, which governs dynamics of the spatial profile of nuclear polarization.
The extremely long oscillation periods (up to hundreds of seconds) observed in experiments as well as the differences in phenomenology between vertical and lateral quantum dot structures are naturally explained in the proposed framework.
 
\end{abstract} 
 
\maketitle 

The coupling of electron and nuclear spin dynamics is responsible for a wide variety of intriguing transport phenomena in semiconductor devices.
Spin exchange between electron and nuclear spins provides a mechanism for electron spin flips which can dramatically alter the behavior of systems such as spin-blockaded quantum dots, where transport is highly sensitive to spin selection rules\cite{OnoSB, Hanson07, Coish05, Jouravlev06, Klauser07, Qassemi09, Danon09}.
Furthermore, the nuclear spins produce a hyperfine (Overhauser) field that shifts the electronic Zeeman energy by an amount corresponding to an effective field that may reach as high as a few Tesla when the nuclei are fully polarized.
This Overhauser field can have dramatic consequences for transport in quantum dots, where discrete levels may be shifted in-to or out-of resonance\cite{OnoTarucha, Jouravlev06, Koppens, Baugh}.
The combination of these two effects -- electron-nuclear spin-exchange which polarizes nuclear spins, and subsequent back-action on energy-dependent spin flip rates -- 
is responsible for a variety of interesting nonlinear dynamical effects such as multistability, hysteresis, and intermittency\cite{OnoTarucha, Koppens, Baugh, RudnerDNP, Inarrea07, Danon07, Vink09, Fluctuations}. 

Among all of the nonlinear phenomena which have been observed thus far in transport through double quantum dots (DQDs), perhaps the most striking is the appearance of spontaneous, stable current oscillations under the application of a dc source-drain bias\cite{OnoTarucha, Austing07}.
This phenomenon is remarkable for a number of reasons.
First, the oscillations occur with very long periods ranging from seconds to hundreds of seconds.
These timescales are $10^7 - 10^9$ times longer than the $(1\ {\rm pA})/e \sim 100$ ns microscopic timescale associated with single electron tunneling through the double dot. 
Second, the oscillations are accompanied by long transients and long memory times when the source-drain bias is switched off and on.
Finally, after many years of experiments by a variety of groups, the oscillations have only ever been seen in {\it vertical} DQDs; the phenomenon has never been observed in a gate-defined lateral double quantum dot.

\begin{figure}[h]
\includegraphics[width=3.2in]{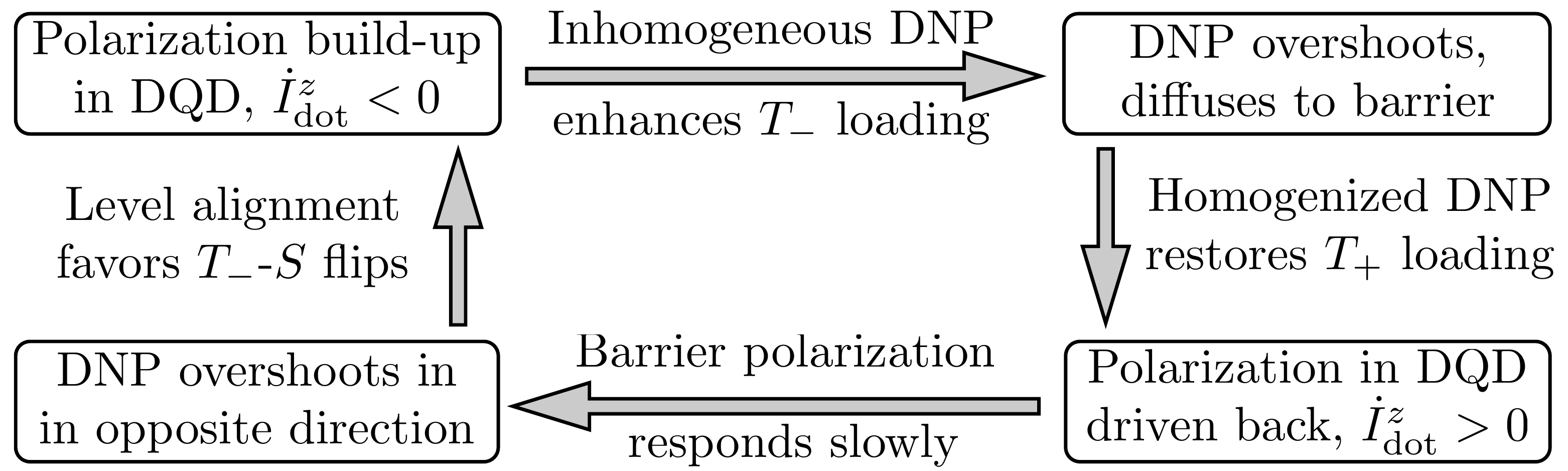} 
 \caption[]{
Mechanism of nuclear polarization oscillations in a spin-blockaded double quantum dot.
Polarization is driven on a short time scale by resonant hyperfine transitions inside the DQD.
Spin injection in the presence of an inhomogeneous Overhauser field leads to a polarization overshoot in the dot.
Nuclear spin diffusion homogenizes the Overhauser field on a much longer timescale.
Spin-flip transition rates inside the DQD adapt but lead to an overshoot in the opposite direction. 
 }
\vspace{-5mm}
\label{figFlowChart}
\end{figure}

Using nuclear magnetic resonance it was shown that the oscillations are in some way driven by nuclear spin dynamics\cite{OnoTarucha}.
However, despite wide interest in the problem,
a viable mechanism has thus far remained elusive.
Here we present a straightforward mechanism which naturally produces oscillations with similar phenomenology. 
The mechanism relies on nuclear spin diffusion \cite{Paget, ReillyDiffusion} and on spin-dependent tunneling rates \cite{Stano10} which are controlled by the spatial profile of the Overhauser field.

Nuclear spin diffusion, being a slow process, introduces the correct timescale into the dynamics. 
Furthermore, it also accounts for a sharp difference in predicted phenomenology for vertical and lateral DQD structures. 
The length scale for out-of-dot diffusion is set by the combination of barrier and quantum well half-widths, and is typically a few tens of nanometers.
For typical diffusion parameters\cite{Paget, ReillyDiffusion} this translates into diffusion times on the order of 10 seconds, consistent with the observed oscillation period values. 
These timescales are much longer than those arising from coherent mechanisms\cite{JouravlevOscillation}.
Additionally, in vertical DQDs such as those used in Ref.~\cite{OnoTarucha,Austing07}, the edges of the dot are defined by the mesa structure itself. 
In such structures, nuclear polarization predominantly diffuses in the vertical direction, 
into the adjacent tunnel barriers.
For gate-defined lateral DQDs, however, nuclear polarization can diffuse in all directions\footnote{In fact diffusion is fastest in the vertical direction due to anisotropy of the quantum dot, thus making transfer to the barrier region even more unlikely.}, and is not expected to flow significantly into the tunnel barriers. 
Thus the feedback effect via polarization-dependent tunneling, which is responsible for the oscillations, is expected to occur in vertical DQDs but not in lateral DQDs. 
This is consistent with the observation that oscillations are frequently observed in vertical structures but never in the lateral structures.

Schematically, oscillations arise as described in Fig.~\ref{figFlowChart}.
An initial imbalance of hyperfine spin flip rates for up and down electron spins leads to a fast build up of nuclear polarization inside the DQD.
The resulting inhomogeneity of the Overhauser field between the DQD and its surroundings {\it enhances} the probability of injecting the spins with the dominant hyperfine rate.
This causes the polarization inside the dot to ``overshoot.'' 
On a much longer time scale, nuclear spin diffusion homogenizes the Overhauser field. 
As the spin-injection probabilities react accordingly, the balance of hyperfine transition rates inside the DQD reverses, and starts to drive the nuclear polarization in the dot back toward zero.
In a similar way, the polarization inside the dot again overshoots and then the cycle repeats.

Here we describe the coupled electron and nuclear spin dynamics through the simplest possible model which qualitatively captures the behavior of the essential physical degrees of freedom.
In principle, detailed numerical modeling of the dot and of the full spatial profile of nuclear polarization could be attempted.
However, such modeling would introduce a much higher level of complexity, and would likely cloud rather than clarify the essence of the oscillation mechanism.
Instead, we will write a set of dynamical equations for two polarization variables, one representing the polarization within the double quantum dot, and one representing the polarization under the tunnel barrier to the source lead.
The intradot polarization variable is driven by hyperfine spin flip processes with electron spins within the DQD.
Polarization is then transferred to the barrier via spin diffusion with a large time constant.
The delayed reaction of the barrier polarization variable to the intradot spin dynamics leads to oscillations as outlined above. 

\begin{figure}[t]
\includegraphics[width=2.8in]{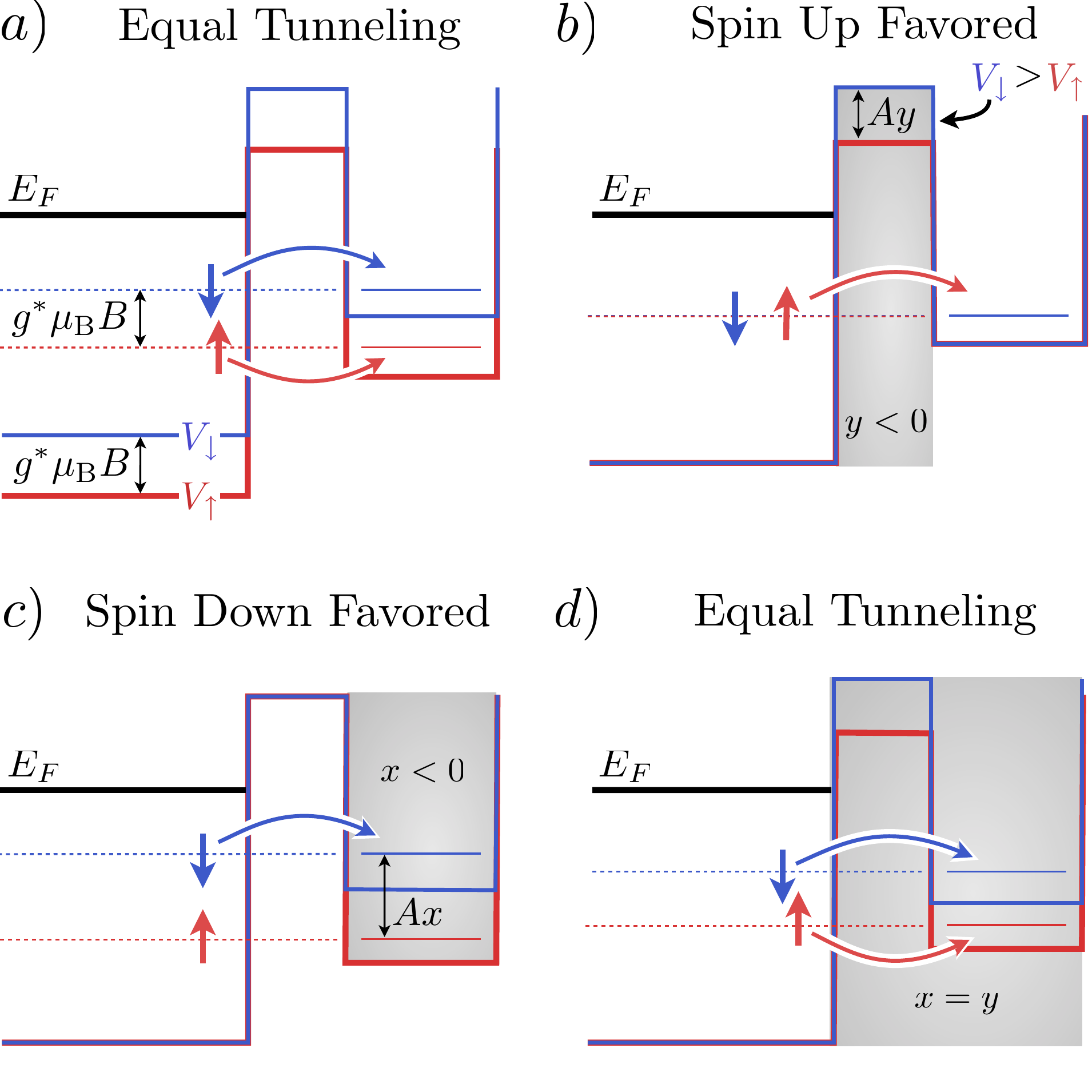} 
 \caption[]{
Spin-dependent tunneling due to inhomogeneous Overhauser field.
a) For a homogeneous Zeeman field, up and down spins are subjected to identical barriers and tunnel into an empty dot with equal probabilities.
b) When nuclear polarization is nonzero only under the barrier, $y \neq 0$, up and down spins are subjected to different barriers ($B = 0$ for illustration in b-d). 
c) When nuclear polarization is nonzero only inside the quantum dot, $x \neq 0$, up and down spins tunnel in at different relative energies. 
d) If the nuclear polarizations in the dot and in the barrier are nonzero but equal, $x = y$, up and down spins tunnel in with equal probabilities.
 }
\vspace{-7mm}
\label{fig1}
\end{figure}

A key to our mechanism is the difference in probabilities for spin-up and spin-down electrons to tunnel into the quantum dot when it is empty\cite{Stano10}. 
Naively, one might expect the respective tunneling rates to differ due to the application of a {\it homogeneous} Zeeman field, since the final state energies are different.
However, as shown in Fig.\ref{fig1}a, up and down spins 
tunnel under identical Zeeman-shifted barriers.
Provided that the dot levels are set far below the chemical potential of the lead, $E_F$, and that the lead has an approximately constant density of states in the energy range of interest, the tunneling-rates for the two spin species are equal in the case of a homogeneous field.

What happens in the case of an inhomogeneous Zeeman-Overhauser field?
For demonstration, consider the case shown in Fig.\ref{fig1}b, where the nuclear polarization is large under the barrier and zero outside.
Here the Overhauser field locally {\it increases} the Zeeman energy under the barrier, effectively creating a higher barrier for down spins, and a lower barrier for up spins.
In this situation, an empty dot is more likely to be filled by a spin-up electron than by a spin-down electron.
Similarly, as shown in Fig.\ref{fig1}c, nuclear polarization concentrated only inside the dot can also affect the tunneling-in probabilities by changing the tunneling energies relative to the tops of the spin-up and spin-down barriers.

It is interesting to note the similarity between this effect and the phenomenon of tunneling
magnetoresistance (TMR)\cite{Moodera,Miyazaki}. In both cases transport is dominated by
tunneling through a barrier and spin polarization is used as a knob to control
tunneling rate. While in TMR the spin polarization is due to magnetization in
the regions surrounding the barrier, in our case the dominant effect is due to
under-barrier nuclear spin polarization. The discovery of TMR has had important consequences for magnetic memory applications.
One can envision that some of these ideas can be transposed to DQD systems.

\begin{figure}[t]
\includegraphics[width=3.0in]{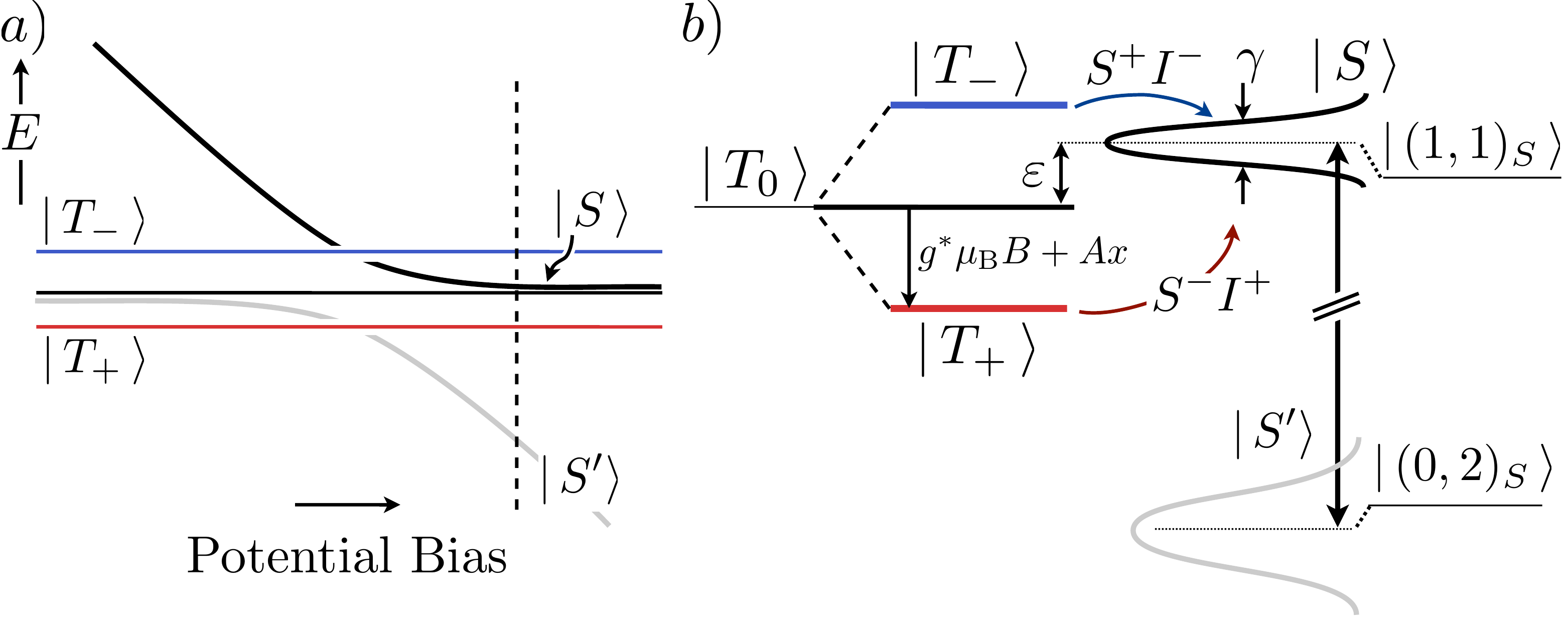} 
 \caption[]{
Two-electron energy levels involved in spin-blockaded transport (adapted from Ref.\cite{AsymmB}).
a) As a function of interdot potential bias, which controls the asymmetry of the double well potential, the (1,1) and (0,2) singlet states exhibit an anticrossing.
b) Energy levels at large detuning, indicated by the dashed vertical line in a).
The singlet levels are broadened due to the coupling of the (0,2) state to the drain lead.
Hyperfine-assisted transitions from $\Ket{T_\pm}$ to $\Ket{S}$ provide a source for the nuclear polarization $x$ within the double dot.
 }
\vspace{-7mm}
\label{LevelDiagram}
\end{figure}

We now consider sequential electron transport through a spin-blockaded double quantum dot connected to leads with an applied dc source-drain bias, as described for example in Refs.~\cite{OnoTarucha, RudnerDNP, Inarrea07}. 
In the two-electron spin-blockade regime, ``(1,1)'' orbital configurations with one electron in each dot, and a ``(0,2)'' configuration with both electrons in the second dot have nearly the same electrostatic energies.
In the (1,1) configuration, where overlap between electrons is negligible, all four spin states (one singlet and three triplet states) are nearly degenerate in energy.
For the (0,2) configuration, however, only the spin singlet configuration is allowed due to the Pauli exclusion principle (the single dot orbital level spacing is assumed to be much larger than the applied bias).
Interdot tunneling hybridizes the (1,1) and (0,2) singlet states, producing the states labeled $\Ket{S}$ and $\Ket{S'}$ in Fig.~\ref{LevelDiagram}.

Tunneling out of the double dot occurs from the (0,2) singlet state, which is coupled to the drain lead.
Through hybridization, both singlet states $\Ket{S}$ and $\Ket{S'}$ acquire finite lifetimes, reflected in their broadened lineshapes as shown in Fig.~\ref{LevelDiagram}b.
When only spin-conserving tunneling processes are taken into account, the triplet states remain decoupled from the drain.
Therefore the rate-limiting step which controls current through this system is the decay of the long-lived triplet states through resonant hyperfine-assisted transitions to the singlet states $\Ket{S}$ and $\Ket{S'}$, or higher order processes which may also break the conservation of spin within the double dot\cite{Qassemi09}.
Hyperfine assisted transitions from $\Ket{T_\pm}$ to $\Ket{S}$ and $\Ket{S'}$ transfer angular momentum from electron to nuclear spins, and thus drive the nuclear polarization dynamics.

Here we focus on the regime of large detuning where the level $\Ket{S'}$ is far separated in energy from the triplet states and can be ignored in the calculation of hyperfine-assisted triplet-singlet transitions.
We seek a coupled set of dynamical equations in two polarization variables.
The first variable, $x = (N_+ - N_-)/(N_+ + N_-)$, represents the fractional polarization inside the double dot, where $N_+$ ($N_-$) is the number of nuclear spins oriented along (against) the external field.  
For a typical device, $N \equiv N_+ + N_- \approx 10^6$.
The second variable, which we denote by $y$, represents the fractional polarization within the tunnel barrier connecting the source lead to the first dot.

The intradot polarization $x$ controls feedback through the Overhauser shift of the electronic triplet levels, which can bring these levels into or out-of resonance with the singlet.
The energies $\epsilon_\pm$ of the triplet states $\Ket{T_\pm}$, relative to $\Ket{S}$, are given by
\begin{equation}
\epsilon_\pm = \epsilon \pm g^*\mu_{\rm B} B \pm Ax,
\end{equation}
where $\epsilon$ is the singlet-triplet detuning, $g^*$ is the electronic effective g-factor ($g^* \approx -0.44$ in GaAs), $\mu_{\rm B}$ is the Bohr magneton, $B$ is the strength of the applied magnetic field, and $A \sim 100\ \mu{\rm eV}$ is the hyperfine coupling constant.

Each time an electron decays from $\Ket{T_+}$ or $\Ket{T_-}$ to $\Ket{S}$ via hyperfine exchange, one nuclear spin is flipped from down to up, or up to down, respectively.
The probability for an electron that enters the dot to cause a positive (negative) increment to the nuclear polarization during its escape is determined by the probability $f_+$ ($f_-$) that the electron entered into the state $\Ket{T_+}$ ($\Ket{T_-}$), and by the probability that the electron escapes via the hyperfine exchange process 
rather than by alternative nuclear-spin-independent mechanisms\cite{RudnerDNP,AsymmB}. 
The hyperfine spin flip probabilities are determined by the ratios $W^{\rm HF}_\pm/(W^{\rm HF}_\pm + W^{\rm in})$, where $W^{\rm HF}_\pm$ is the hyperfine decay rate of $\Ket{T_\pm}$ and  $W^{\rm in}$ describes the collective effects of spin-orbit coupling, spin exchange with the leads, and cotunneling processes.

In our model, we assume that all nuclear spin flips due to hyperfine exchange with the electrons occur {\it within} the double quantum dot.
Therefore the dot polarization $x$ receives kicks (with magnitude $1/N$) on the timescale of single electron hopping through the dot, 100 ns to 1 $\mu$s, while the barrier polarization $y$ has no dynamics on this small timescale. 
On a much longer timescale, nuclear polarization may diffuse from the dot region into the barrier region, providing a source for $y$.

Mathematically, it is simplest to analyze the regime where $W^{\rm in} \gg W_\pm^{\rm HF}$.
Here the total current, i.e.~the effective frequency of electrons passing through the double dot, is determined by $W^{\rm in}$.
Additionally, the hyperfine decay probabilities reduce to $W^{\rm HF}_\pm/W^{\rm in}$.
The dependence on $W^{\rm in}$ cancels from the nuclear polarization {\it rate}, which depends on products of attempt frequencies and spin flip probabilities, leaving behind contributions proportional to the hyperfine rates $W^{\rm HF}_\pm$ weighted by the loading probabilities $f_\pm$:
\begin{eqnarray}
\label{xdot}\dot{x} &=& f_+ W^{\rm HF}_+\ -\ f_-W^{\rm HF}_-\ -\ 2\Gamma_D(x - y)\\
\label{ydot}\dot{y} &=& -2\Gamma_Dy  + \Gamma_Dx,
\end{eqnarray}
where $\Gamma_D \sim 0.1\ s^{-1}$ is the inverse of the time constant for diffusion from the dot to the barrier.
The hyperfine spin-flip rates $W^{\rm HF}_\pm$ are given by Fermi's golden rule\cite{RudnerDNP}:
\begin{equation}
W^{\rm HF}_\pm = \frac{A^2}{N}\frac{(1-x)\,\gamma}{\epsilon_\pm^2 + \gamma^2},
\end{equation}
where $\gamma$ is the decay rate of $\Ket{S}$ due to its coupling to the drain.
We account for the dependence of the loading probabilities $f_\pm$ on the Overhauser field inhomogeneity in a lowest-order expansion in $x$ and $y$:
\begin{equation}
\label{fpm}f_\pm = \frac14[1\pm \eta (x-y)],
\end{equation}
where $\eta$ controls the sensitivity of the loading probabilities to a polarization gradient.
The factors of 2 in front of $\Gamma_D$ in Eqs.~(\ref{xdot}) and (\ref{ydot}) account for the fact that polarization diffuses in both directions (up and down).

Under what conditions might we expect to find oscillations in the flow defined by Eqs.~(\ref{xdot}) and (\ref{ydot})?
Typically, oscillations are found when the linearized system has the form of an ``unstable spiral:''
\be\label{eq:linear_system}
\dot{u} = \alpha u + v,\quad
\dot{v} = -\mu u + \beta v,
\ee
with $(\alpha + \beta) > 0$ and $(\alpha - \beta)^2 - 4\mu < 0$.
These conditions ensure that the eigenvalues are complex, with positive real part.
Comparing with Eq.(\ref{ydot}), we see that $\dot{y} \sim x$, with a {\it positive} coefficient of $x$ due to the fact that polarization preserves its sign as it flows into the barrier. 
Therefore, 
we need the coefficient of $y$ in Eq.(\ref{xdot}) to be {\it negative}.
Substituting expression (\ref{fpm}) for $f_\pm$ into Eq.(\ref{xdot}), we thus obtain a threshold $\eta > 4\Gamma_D/W_0$, where $W_0$ is the hyperfine spin flip rate at the unstable fixed point. 

\begin{figure}[t]
\includegraphics[width=3.0in]{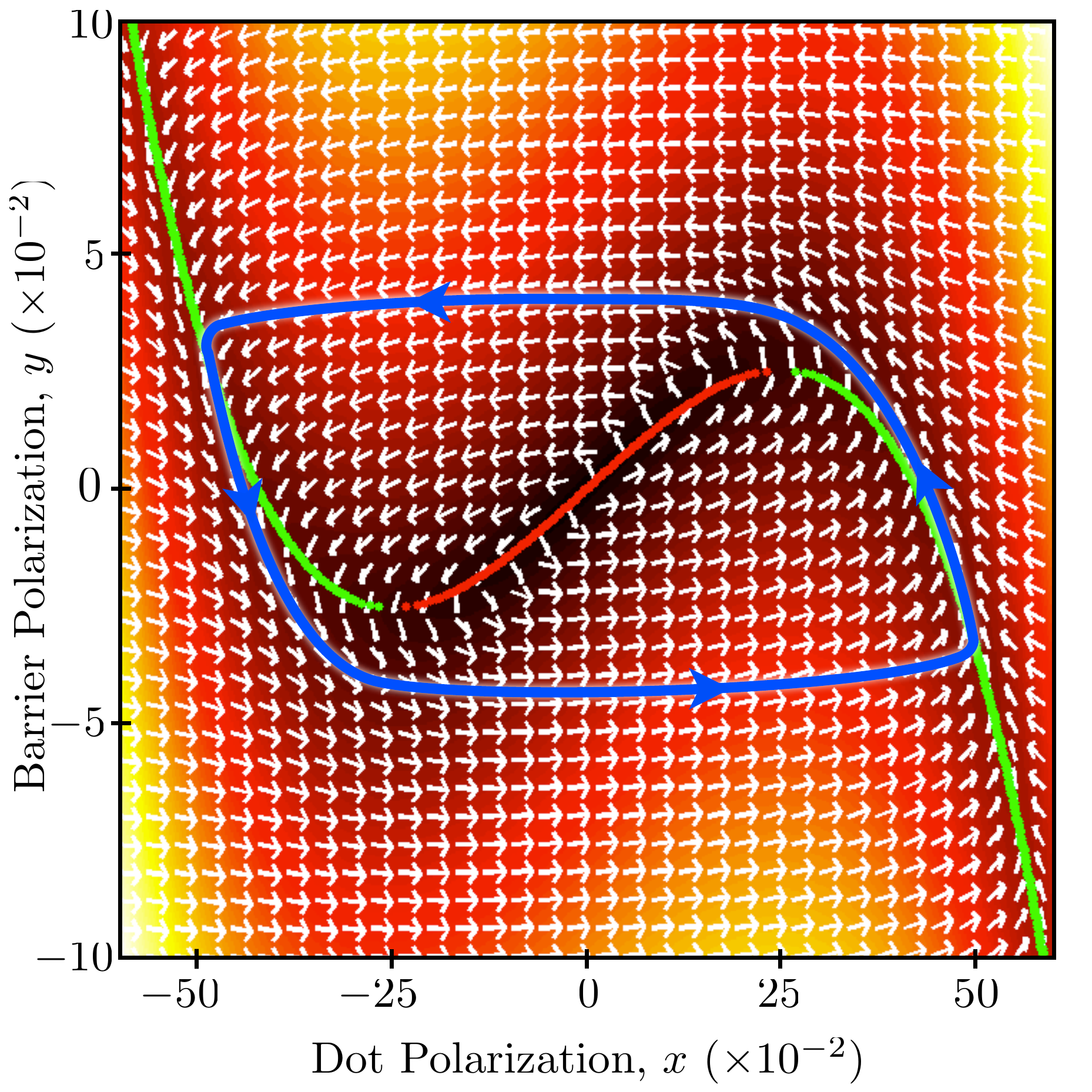} 
 \caption[]{
Polarization velocity field in the oscillatory regime.
Parameter values: $\epsilon/A = -3$, $\gamma/A = 0.05$, $B = 0$, $\eta = 0.4$, $\Gamma_D/A = 10^{-11}$.
Arrows indicate the direction of the velocity field $(\dot{x},\dot{y})$, while the color scale indicates the magnitude of the velocity (arbitrary units).
The green and red curves indicate branches of stable and unstable fixed points of the quasi-one-dimensional dynamics (\ref{xdot}) with $y$ held constant.
The blue curve shows the approximate trajectory of the limit cycle.
 }
\vspace{-7mm}
\label{figVField}
\end{figure}

Going one step further, we can expand Eqs.~(\ref{xdot}) and (\ref{ydot}) 
in the deviations $\tilde{x}$ and $\tilde{y}$ from the (unstable) fixed point of the nonlinear system.
Notably, because $y$ only appears to linear order in the original expressions, only $\tilde{y}$-independent or $\tilde{y}$-linear terms will show up in the expansion.
In general, all other terms will appear: 
\be
\dot{\tilde{x}} \approx  c_{10} \tilde{x} + c_{01} \tilde{y} + \cdots
,\quad
 \quad \dot{\tilde{y}} = \Gamma_D\tilde{x} - 2\Gamma_D\tilde{y}.
\ee
where the dots 
represent higher order terms $c_{20} \tilde{x}^2 + c_{11} \tilde{x}\tilde{y} + c_{30} \tilde{x}^3 + c_{21}\tilde{x}^2\tilde{y} + \cdots$.
Comparing to Eq.(\ref{eq:linear_system}), we see that
we need $c_{10} > 2\Gamma_D > 0$ to ensure a positive real part of the eigenvalues, and $c_{01} < 0$, $(c_{10} + 2\Gamma_D)^2 < 4|c_{01}|\Gamma_D$ to ensure a negative discriminant. 
These considerations 
lead to the oscillatory regime shown in Fig.~\ref{figVField}. 

Using the vast separation of timescales between the hyperfine spin-flip driven polarization dynamics and the slow diffusion processes, we explore another avenue of analysis.
We assume that the barrier polarization $y$ is constant on the timescale of changes in the dot polarization and examine the fixed points of the resulting quasi-one-dimensional dynamical system (\ref{xdot}). 
Figure \ref{figVField} shows the corresponding stable (green) and unstable (red) fixed points, superimposed on the velocity field map of the full system (arrows indicate direction of the polarization velocity, and the color scale indicates its magnitude).
The oscillations can be seen as resulting from the combination of the circulatory flow around the origin, ensured by the spiral condition above, combined with the existence of the unstable branch of quasi-fixed-points near the origin.
Fast horizontal motion towards the quasi-stable points, followed by slow drift along the green curves provides a pictorial representation of the oscillation mechanism outlined in Fig.~\ref{figFlowChart}.
The oscillation period is dominated by the length of the excursions along the green branches.
As a result, the period grows as the oscillation amplitude increases, consistent with experiment\cite{OnoTarucha}.
As parameters vary, a transition out of the oscillation regime occurs when the unstable branch shrinks and disappears. 

We have identified a straightforward physical mechanism which can produce stable oscillations of dynamical nuclear polarization in spin-blockaded vertical double quantum dots.
The mechanism relies on nuclear spin diffusion into a tunnel barrier and is active only in vertical DQDs.
The dependence of spin-injection probabilities and spin diffusion times on barrier width provides a clear experimental signature of this mechanism.
Persistent oscillations can serve as a new probe of nuclear spin diffusion and spin dynamics in vertical structures.

We gratefully acknowledge helpful discussions with S. Amaha, D. G. Austing, and S. Tarucha.
M. R. thanks the Institute for Quantum Optics and Quantum Information for their generous hospitality and support.


\end{document}